\documentclass[journal,comsoc]{IEEEtran}
\usepackage[T1]{fontenc}% optional T1 font encoding

\usepackage{cite}
\usepackage[pdftex]{graphicx}
\graphicspath{{Figures/}}

\usepackage{amsmath}
\usepackage[cmintegrals]{newtxmath}
\usepackage{enumitem}
\setlist[description]{wide=\parindent}
\usepackage{algorithmic}
\usepackage{algorithm}
\usepackage{array}
\usepackage[caption=false,font=normalsize,labelfont=sf,textfont=sf]{subfig}

\usepackage[dvipsnames]{xcolor}
\begin{document}
\title{Geometric Constellation Shaping for Fiber-Optic Channels via End-to-End Learning}

\author{Ognjen~Jovanovic,~\IEEEmembership{Member,~IEEE,}
        Francesco~Da~Ros,~\IEEEmembership{Senior Member,~IEEE,}
        Darko~Zibar,~%\IEEEmembership{}% <-this % stops a space
        and~Metodi~P.~Yankov,~\IEEEmembership{Member,~IEEE}
\thanks{O. Jovanovic, F. Da Ros, D. Zibar, and M. P. Yankov are with the Department
of Electrical and Photonic Engineering, Technical University of Denmark, 2800 Kgs. Lyngby,
Denmark, e-mail: ognjo@dtu.dk}% <-this % stops a space
}


\markboth{ECOC Invited}%
{Preprint}


\IEEEspecialpapernotice{(Invited Paper)}

\maketitle

\begin{abstract}
End-to-end learning has become a popular method to optimize a constellation shape of a communication system. When the channel model is differentiable, end-to-end learning can be applied with conventional backpropagation algorithm for optimization of the shape. A variety of optimization algorithms have also been developed for end-to-end learning over a non-differentiable channel model. In this paper, we compare a gradient-free optimization method based on the cubature Kalman filter, model-free optimization and backpropagation for end-to-end learning on a fiber-optic channel modeled by the split-step Fourier method. The results indicate that the gradient-free optimization algorithms provide a decent replacement to backpropagation in terms of performance at the expense of computational complexity. Furthermore, the quantization problem of finite bit resolution of the digital-to-analog and analog-to-digital converters is addressed and its impact on geometrically shaped constellations is analysed. Here, the results show that when optimizing a constellation with respect to mutual information, a minimum number of quantization levels is required to achieve shaping gain. For generalized mutual information, the gain is maintained throughout all of the considered quantization levels. Also, the results imply that the autoencoder can adapt the constellation size to the given channel conditions.
\end{abstract}


% \fi
\IEEEpeerreviewmaketitle

\section{Introduction}
\IEEEPARstart{O}{ptical} communication systems need to increase their throughput in order to keep up with the growth of data traffic demand. In the linear region, optimization of the modulation formats using geometric or probabilistic constellation shaping can achieve capacity approaching throughput. Probabilistic constellation shaping (PCS) is a well established approach for increasing the throughput and providing rate adaptivity \cite{Bocherer_TCOM15}. However, it comes at the expense of increased transceiver complexity because it requires a distribution matcher and dematcher. Due to the serial nature of the distribution matcher, hardware efficient implementation could prove to be challenging for future systems in which a single carrier will transmit data rates above 1Tb/s \cite{Millar2019}. Also, the interplay of PCS and digital signal processing (DSP) can lead to deteriorated performance, especially in the case of the phase and frequency recovery \cite{barbosa2019phase}. In \cite{Mello_JLT2018}, it was demonstrated that for moderate to low signal-to-noise ratio (SNR) the quality of the phase estimation of the blind phase search (BPS) algorithm\cite{Pfau2009a} is degraded in the presence of PCS. Geometric constellation shaping (GCS) could prove to be a low-complexity alternative to PCS because it is directly compatible with the classical bit-interleaved coded modulation (BICM) and can be optimized with respect to existing DSP \cite{Jones2019}. However, it should still be mentioned that some low-complexity PCS algorithms have found their way in commercial DSP implementations, e.g. \cite{Jannu}.

End-to-end learning, which was introduced in \cite{OShea2017}, utilizes an autoencoder (AE) \cite{goodfellow2016deep} to optimize the communication system and it has gathered traction as a method to approach GCS due to its versatility to the employed channel model. End-to-end learning has proven to be effective for GCS in optical communication systems \cite{Jones2018a,Jones2019,Li2018b,Schaedler2020,Gumus2020,Vlad_ECOC,Oliveira:22}, mainly focusing on the mitigation of the nonlinear effects of the optical fiber. Geometric constellation shaping considering the BPS algorithm was explored in \cite{Ognjen_ECOC,Ognjen_ECOC_Extension}, where a constellation was optimized to be robust to channel uncertainties with BPS at the receiver. Joint probabilistic and geometric constellation shaping can also be approached with end-to-end learning \cite{Aref2022,Neskorniuk:22}. Apart from constellation shaping, end-to-end learning was applied in optical communication for waveform optimization for dispersive fiber \cite{Karanov2018,Karanov2019a,Karanov2021}, waveform optimization for nonlinear frequency division multiplexing \cite{Gaiarin2020,GaiarinJLT} and superchannel transmission \cite{Song:21,Song2022}.

Typically, an AE is optimized using backpropgation (BP) which relies on a gradient-based algorithm and it requires that the embedded channel model is differentiable. All the aforementioned work fulfilled this requirement and performed gradient-based optimization. However, a differentiable channel might not always be available which makes this requirement too strict. One example is non-numerical channels such as experimental test-beds. Another example is channels including decision-directed DSP algorithms such as the classical BPS algorithm and decision-directed equalizers. Alternatives have been proposed, such as modeling these channels using generative adversarial networks \cite{o2019approximating,ye2018channel,Karanov2020}, avoiding to propagate the gradient through the channel \cite{Model-free, Hoydis, Ognjen_JLT} or adapting the channels to be differentiable \cite{rode2021geometric}. 

In \cite{Model-free,Hoydis}, a two-stage alternating algorithm which relies on reinforcement learning (RL) was used to train the AE without a known channel model. This approach was utilized to train a NN for digital predistortion over-the-fiber \cite{Song_ecoc2021}. Optimization of an AE with a non-differentiable channel, more specifically the BPS algorithm, using a gradient-free optimization was proposed and investigated in \cite{Ognjen_JLT}. A differentiable version of the BPS was proposed in \cite{rode2021geometric} in order to optimize the AE using classical gradient-based algorithms. It was expanded to include robustness to channel uncertainties \cite{rode2022ECOC} similar to \cite{Ognjen_ECOC}.

The AE-based GCS does not require hardware changes as it simply optimizes the positions of the constellation points in the complex plane and can be implemented as a lookup table \cite{Jones2019}. However, the geometrically shaped constellation points may require a finer quantization than a typical square quadrature amplitude modulation (QAM), which could prove to be the main drawback of GCS \cite{Qu2019}. End-to-end learning of a communication system which includes quantization noise was done e.g. in \cite{Karanov2018,Song:21}. However, a fixed number of quantization bits was used and the effect of different number of quantization bits was not discussed.

This paper is an extension of \cite{Metodi_ECOC}, which compares different AE training algorithms with respect to their performance and analyses the impact of quantization noise when optimizing a constellation with respect to mutual information (MI). In this paper, the results from \cite{Metodi_ECOC} are discussed and the analysis of the training algorithms is extended to include convergence and complexity. The impact of quantization noise is then extended to an optimization of a constellation with respect to generalized mutual information (GMI) and compared to the ones obtained with MI. An analysis of the differences between the constellations optimized for MI and GMI is included.

The remainder of the paper is organized as follows. MI and GMI are used as performance metrics and the basic principles of estimating them are described in Section \ref{PM}, as well as the principle of mismatched decoding. In Section \ref{sec:AE_GCS}, the AE-based GCS with respect to MI and GMI is described, as well as the applied optimization algorithms. A detailed description of the system, the AE architecture, channel models and the optimization procedure, are provided in Section \ref{sec:systems}. Section \ref{results} provides the results on the MI performance of the constellations trained with different optimization algorithm on the split step Fourier method (SSFM) channel model, as well as MI and GMI performances obtained by constellations trained on a channel which includes quantization noise. The conclusions are summarized in Section \ref{Conclusion}.

%\IEEEpubidadjcol

\section{Performance metrics}\label{PM}
\subsection{Mutual information}
Let $\mathcal{X}$ be a set of complex constellation points (symbols) with cardinality $|\mathcal{X}|=M=2^m$, where $m$ is the number of bits carried by a symbol. Consider $\mathbf{B}=[B_1,B_2,\dots,B_m]$ a random variable of binary $m$-dimensional vectors which are mapped to complex valued symbol $X \in \mathcal{X}$ with a uniform probability mass function $P_{X}(x)=\frac{1}{M}$. The channel transition probability density $p_{Y|X}(y|x)$ governs the channel input-output relation, where $Y$ is a complex output of a channel with $X$ as input. The channel output $Y \in \mathbb{C}$ has a probability distribution $p_{Y}(y)$, where $\mathbb{C}$ denotes the set of complex numbers. MI $I(X;Y)$ represents the amount of information shared between $Y$ and $X$ in bits per symbol,
\begin{equation}
\begin{split}
    I(X;Y) & = H(X)-H_p(X|Y) = m - H_p(X|Y) \\
    & = \sum_{x \in \mathcal{X}}P_{X}(x) \int_{\mathbb{C}} p_{Y|X}(y|x)\log_2 \frac{p_{Y|X}(y|x)}{p_{Y}(y)}dy \text{,} 
\end{split}
\label{eq:MI}
\end{equation}
where $H(X)= -\sum_{x \in \mathcal{X}}P_{X}(x)\log_2(P_{X}(x))=\log_2(M)=m$ is the entropy of $X$ and $H_p(X|Y)= \mathbb{E}_{p(x,y)}[p_{X|Y}(x|y)]$ is the conditional entropy of $X$ given $Y$. The probability $p_{X|Y}(x|y)$ is the posterior probability of $X$ given $Y$. The expectation $\mathbb{E}_{p(x,y)}$ should be taken over the true joint probability density function of $p_{X,Y}(x,y)$.

In order to evaluate Eq. (\ref{eq:MI}), it is required to know the channel transition probability $p_{Y|X}(y|x)$ and its analytical expression, however, in optical communication this is often not the case. When the transition probability $p_{Y|X}(y|x)$ is unknown or the analytical expression is unavailable, a typical approach is to bound Eq. (\ref{eq:MI}). Instead of the true transition probability $p_{Y|X}(y|x)$, a transition probability $q_{Y|X}(y|x)$ of an auxiliary channel is considered \cite{Lowerbound}. This approach is known as mismatched decoding and it can be used to obtain a lower bound on the MI which is also known as the achievable information rate (AIR). This lower bound is formulated as 
\begin{equation}
I(X;Y) \geq H(X)-\hat{H}_q(X|Y) = m - \hat{H}_q(X|Y)\text{,}
\label{eq:MI_lower}
\end{equation}
where $\hat{H}_q(X|Y)=\mathbb{E}_{p(x,y)}[q_{X|Y}(x|y)]$ is an upper bound of the true conditional entropy $H_p(X|Y)$. The probability distribution $q_{X|Y}(x|y)$ is the auxiliary distribution of the true posterior probability $p_{X|Y}(x|y)$. The inequality in Eq. (\ref{eq:MI_lower}) turns to equality when $q_{X|Y}(x|y)=p_{X|Y}(x|y)$.

\subsection{Generalized mutual information}

For optical communication, a bit-interleaved coded modulation (BICM) architecture is usually used and it contains a bit-wise demapper. Achieving the previously described MI requires a symbol-wise forward error correction (FEC) or iterative demapping and decoding. Instead, GMI is a good performance indicator for an architecture that employs binary soft-decision FEC, such as the more conventional BICM \cite{Alvarado:15}. Using the conditional entropy $\hat{H}_q(B_i|Y)=\mathbb{E}_{p(x,y)}[q_{B_i|Y}(b_i|y)]$ of the bit $B_i$ at the $i$-th position and the channel output $Y$ instead of $\hat{H}_q(X|Y)$, a lower bound on the GMI $I(\mathbf{B};Y)$ can be defined as \cite{Alvarado:15}
\begin{equation}
\begin{split}
    I(\mathbf{B};Y) & \ge \sum_{i=1}^m I(B_i;Y) \ge H(x) - \sum_{i=1}^m \hat{H}_q(B_i|Y) \\
    & = m - \sum_{i=1}^m \hat{H}_q(B_i|Y) \\
    %&=\sum_{i=1}^m \sum_{x \in \mathcal{X}}P_{X}(x) \int_{\mathbb{C}} p_{Y|X}(y|x)\log_2 \frac{q_{Y|B_i}(y|b_i)}{q_{Y}(y)}dy \text{.} 
\end{split}
\label{eq:GMI}
\end{equation}
The GMI and its lower bound are heavily reliant on the bit-to-symbol labeling, which directly influences the tightness of the first bound in Eq.~(\ref{eq:GMI}). For ideal, binary-reflected Gray labeling, that inequality turns into equality. In practice, some loss can be expected, especially for a poorly designed labeling scheme.

\begin{figure*}[htb] %was better for now
\centering
\includegraphics[width=\textwidth]{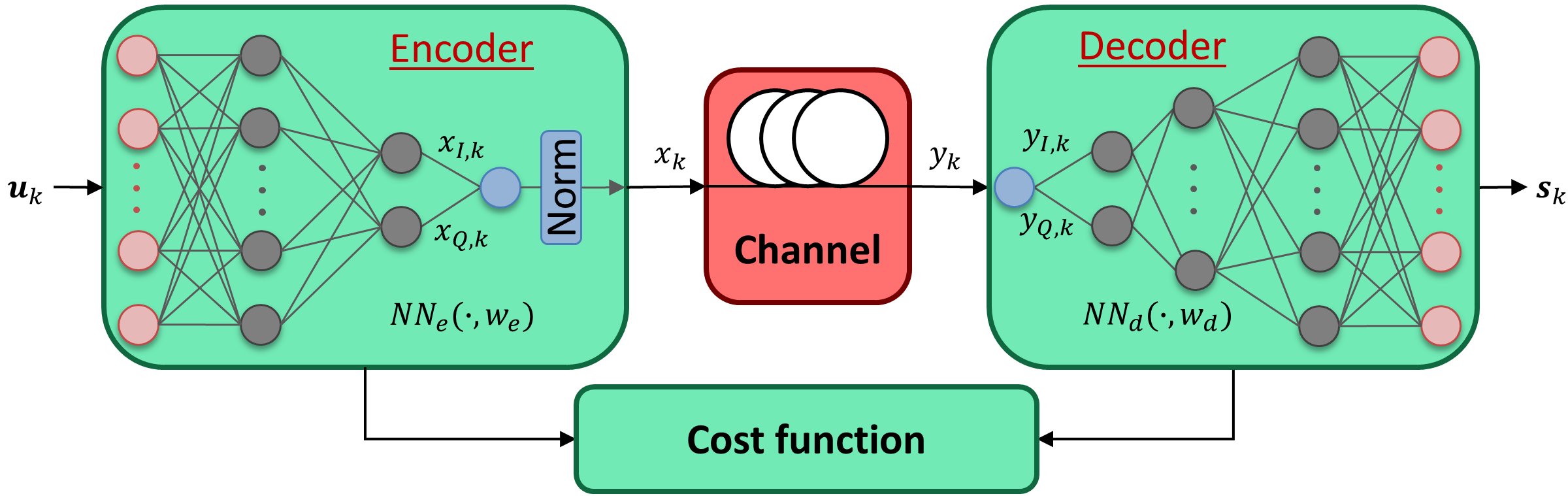}
\caption{Example of autoencoder model for geometrical constellation shaping.}
\label{fig:AE}
\end{figure*}

\subsection{Mismatched Gaussian receiver}
For optical communication, it is a common approach to use a mismatched Gaussian receiver \cite{Lapidoth1994,Metodi:16} which assumes the transition probability $q_{Y|X}(y|x)$ of an auxiliary Gaussian channel \cite{Alvarado2018}
\begin{equation}
   q_{Y|X}(y|x) = \frac{1}{{\pi \sigma_{G}^2}}\exp{\left(-\frac{|y-x|^2}{\sigma_{G}^2}\right)} \text{,}
\end{equation}
where $\sigma_{G}^2$ is the estimated noise variance of the auxiliary channel. The noise variance $\sigma_{G}^2$ may be estimated from the training input-output channel pairs as $\sigma_{G}^2=\mathbb{E}[|y-x|^2]$. By applying the Bayes' theorem, the posterior distributions in Eq. (\ref{eq:MI_lower}) are formulated as
\begin{equation}
    q_{X|Y}(x|y) = \frac{p_X(x)q_{Y|X}(y|x)}{\sum_{\hat{x} \in \mathcal{X}} p_X(x=\hat{x})q_{Y|X}(y|x=\hat{x})}
    \label{eq:qxy}
\end{equation}
and can be evaluated using the Monte Carlo approach. %The auxiliary function $q_{X|Y}(x|y)$ is an approximation to $p_{X|Y}(x|y)$ in two ways: 1) it is modeled using a decoder NN or a Gaussian receiver; 2) it is memoryless. Both of these approximations lead to an upper bound on the conditional entropy and a lower bound on the MI.

\section{Autoencoder-based geometric constellation shaping} \label{sec:AE_GCS}

Geometric constellation shaping may be used to optimize the position of constellation points in high-order modulation formats to improve the throughput and maximize the AIR by either maximizing MI $I(X;Y)$ or GMI $I(\mathbf{B};Y)$. In Fig. \ref{fig:AE}, an example of an AE for GCS is shown and it consists of an encoder neural network (NN), decoder NN and a channel model embedded in between them. Here, feed-forward NNs are used for the encoder and the decoder represented by parametric functions $NN_{e}(\cdot,\mathbf{w}_{e})$ and $NN_{d}(\cdot,\mathbf{w}_{d})$, respectively. The two NNs are parameterized with trainable weights $\mathbf{w}_{e}$ and $\mathbf{w}_{d}$ that also include biases. A weight set that includes both the encoder and decoder trainable weights is defined as $\mathbf{w}=\{\mathbf{w}_e,\mathbf{w}_d\}$ and the total number of the weights is $N_w$. The weights are optimized to minimize the cost function between the input $\mathbf{u}_k$ and the output $\mathbf{s}_k$ of the AE for the given channel model. The cost function is chosen depending on the desired performance metric, e.g. categorical cross-entropy (CE) for MI and binary CE (also known as log-likelihood (LL)) for GMI. For GCS, the encoder learns a constellation, whereas the decoder learns optimal decision boundaries for the learned constellation and considered channel model.

When the desired performance metric is MI, the input to the encoder is a one-hot encoded vector $\mathbf{u}_k \in \mathbb{U} = \{ \mathbf{e}_i | i=1,\dots, M\}$ which is mapped to a normalized complex constellation point $x_k=NN_e(\mathbf{u}_k,\mathbf{w}_{e})$, where $k$ represents the $k$-th sample and $\mathbf{e}_i$ is an all zero vector with a one at position $i$. Here, it is considered that the function $NN_e(\mathbf{u}_k,\mathbf{w}_{e})$ includes the NN which outputs a two-dimensional vector $[x_{I,k},x_{Q,k}]^T$, representing the in-phase and quadrature components of the complex constellation symbol $x_k$, and the normalization of the constellation points $\mathbb{E}_{\mathbf{e}_i}[|NN_e(\mathbf{e}_i,\mathbf{w}_{e})|^2]=1 \quad \text{for} \quad i=1,\dots, M$. The complex symbol is transmitted over the channel, resulting in an impaired symbol $y_k$. The channel can include the fixed blocks of the transceiver, e.g. DSP blocks and hardware components. The in-phase and the quadrature components of the impaired symbol $y_k$ are inputs to the decoder, which outputs a vector of posterior probabilities $\mathbf{s}_k=NN_d(y_k,\mathbf{w}_{d}) \in [0,1]^M$ using a softmax output layer such that $\sum_{i=1}^M\mathbf{s}^{(i)}_k=1$ where $(i)$ denotes the $i$-th element of the output vector. For a simpler notation, the partition of the in-phase and the quadrature components of $y_k$ is considered as a part of the function $NN_d(y_k,\mathbf{w}_{d})$. The trainable AE weights $\mathbf{w}$ are optimized by iteratively minimizing the categorical CE cost function over a sample set of size $N$. In each iteration, the sample set is divided into batches of size $B$ and the CE loss for each batch is calculated as
\begin{equation}
    \begin{split}
    J_{CE}(\mathbf{w}) & = \frac{1}{B} \sum_{k=1}^{B} \bigg[ -\sum_{i=1}^{M} \mathbf{u}^{(i)}_k \log \mathbf{s}^{(i)}_k \bigg].
    \end{split}
    \label{eq:Xent}
\end{equation}
The output of the decoder NN $\mathbf{s}_k$ essentially represents an auxiliary distribution $q_{X|Y}(x|y)$ to the true posterior distribution $p_{X|Y}(x|y)$. Therefore, the CE represents an AE-based upper bound $\hat{H}_{q}(X|Y)=\mathbb{E}_{p(x,y)}[q_{X|Y}(x|y)]$ on the conditional entropy. Based on Eq. (\ref{eq:MI_lower}), this implies that minimizing the CE maximizes a lower bound on the MI and this lower bound is an AIR when using the decoder NN. % is an approximation $q_{X|Y}(x|y)$ of the true posterior distribution $p_{X|Y}(x|y)$.

When the desired performance metric is GMI, three main changes should be made compared to the optimization with respect to MI. First, instead of the one-hot encoded vector as the AE input, a block of $m$ bits $\mathbf{u}_k \in \mathcal{M}=\{0,1\}^m$ should be used, where $\mathcal{M}$ is the set of all possible bit sequences of length $m$. Second, the input and the output space of the AE need to match, therefore the output layer of the decoder should be changed. The sigmoid activation function replaces the softmax function and now the decoder output is $\mathbf{s}_k \in [0,1]^m$ which represents the posterior probabilities of the bits being "0" or "1". Finally, the LL cost function is used because it suits the change of the input/output space,
\begin{equation}
    \begin{split}
        &J_{LL}(\mathbf{w})=\\
        &\frac{1}{B}\sum_{k=1}^{B}\left[-\frac{1}{m}\sum_{i=1}^{m}     \mathbf{u}_k^{(i)}\log{(\mathbf{s}_k^{(i)})}+(1-\mathbf{u}_k^{(i)})\log{(1-\mathbf{s}_k^{(i)})}\right],
    \end{split}
    \label{eq:log-like}
\end{equation}
where $(i)$ denotes the $i$-th element of the $m$--dimensional output. In this case, the decoder output $\mathbf{s}_k$ is an approximation of $q_{B_i|Y}(b_i|y)$ from Eq. (\ref{eq:GMI}). By minimizing the LL cost function, a lower bound on the GMI is maximized in a similar fashion to the maximization of the lower bound on the MI using the categorical CE.

The classical optimization method for an AE is BP of the gradients from the cost function to the trainable weights $\mathbf{w}$. This method requires the embedded channel model to be differentiable and computationally tractable. The former raises an issue with e.g. experimental test-beds and channels which include non-differentiable DSP blocks as aforementioned. The latter raises the practical issue of computational memory and running time of the forward and the backward propagations. An example here is the fiber-optic channel modeled by the SSFM which requires thousands of steps to accurately model a long-haul transmission. It is necessary that each of these steps is stored as a part of the computational graph required by the BP algorithm, making it infeasible for black-box application. 

These drawbacks can be potentially addressed by utilizing optimization strategies that do not require computing the gradient of the channel, such as the RL-based optimization from \cite{Model-free, Hoydis} and gradient-free optimization from \cite{Ognjen_JLT} which utilizes the cubature Kalman filter (CKF) \cite{haykin2009cubature}. In the RL-based optimization, the encoder is trained while the decoder is fixed and vice-versa. The decoder is optimized using the classical BP, whereas the encoder is assumed to be stochastic with a known perturbation and by combining with an RL method a surrogate gradient is calculated for the optimization \cite{Model-free}. The CKF approach relies on describing the AE as a state-space model and estimating the trainable weights using conventional Bayesian inference.

\section{System description and optimization procedure} \label{sec:systems}

In this paper, two separate systems are observed, one for comparison of optimization algorithms and the other to analyse the impact of quantization on GCS.

%\subsection{Autoencoder models}
\subsection{Comparison of optimization algorithms} \label{ssec:SSFM}

\begin{figure}[!t]
    \centering
    \includegraphics[width=\columnwidth]{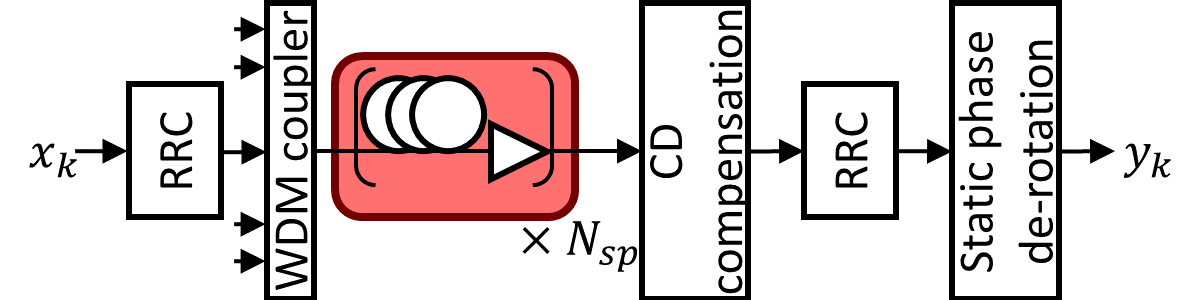}
    \caption{Fiber-optic channel modeled with SSFM for dual polarization, 5 channel WDM transmission with RRC pulse shaping, CD compensation and static phase rotation.}
    \label{fig:comparison_channel}
\end{figure}

The BP, RL-based (further referred to as just RL) and CKF algorithm for optimization of the AE are compared and a comprehensive study of algorithm performance and complexity is performed. For this purpose, a computationally complex channel model, which requires significant amount of computational memory and running time, was embedded into the AE. The channel model is a dual polarization wavelength-division multiplexing (WDM) fiber-optics system and it is shown in Fig. \ref{fig:comparison_channel}. Identical encoder NNs were used to generate symbols for both polarizations of each of the WDM channels. It should be mentioned that an ideal laser and a linear I/Q modulation was assumed. In each channel, the output of the encoder NN is pulse shaped by the root-raised cosine (RRC) filter with a $0.01$ roll-off and the resulting signal is rescaled to the launch power per channel $P_{s}$. These signals are added together in the WDM coupler to form a WDM signal. This signal is transmitted over a fiber-optic channel modeled by the SSFM, simulating a link of $N_{sp}$ EDFA-amplified spans of length $L$. At the receiver, a coherent detection with an ideal local oscillator is considered. Chromatic dispersion (CD) compensation is performed and the central channel is filtered out by a low pass filter which in this case is the RRC matched filter. Static phase de-rotation is performed to finally obtain the signal that will be used as the decoder input. It should be noted that the static phase de-rotation can also be learned by the decoder since it would just have to rotate the decision boundaries. However, one of the goals of this study is to monitor the AIR of the constellation using the mismatched Gaussian receiver, which requires that the transmitted and received constellations are in-phase. This in turn requires a static phase de-rotation to compensate for the average nonlinear phase shift. The parameters of the desired transmission link are given in Table~\ref{tab:channel}.

\begin{table}[!t]
\renewcommand{\arraystretch}{1.5}
\newcolumntype{C}[1]{>{\centering\let\newline\\\arraybackslash\hspace{0pt}}m{#1}}
\caption{Channel parameters}
\label{tab:channel}
\centering
\begin{tabular}{| C{0.45\columnwidth} | C{0.45\columnwidth}|}
\hline
 Symbol rate ($R_s$) & 32 GHz\\
\hline
Carrier frequency ($F_c$) & 193.41 THz\\
\hline
Number of WDM channels & 5\\
\hline
Channel spacing & 50 GHz\\
\hline
Number of polarizations ($N_{pol}$) & 2\\
\hline
Number of spans ($N_{sp}$) & 10\\
\hline
Span length ($L$) & 100 km\\
\hline
Nonlinear coefficient & $1.3$ $(\text{W km})^{-1}$\\
\hline
Dispersion parameter & 16.464 ps/(nm km)\\
\hline
Attenuation ($\alpha$) & 0.2 dB/km\\
\hline
Amplifier gain ($G$) & $\alpha L$\\
\hline
Amplifier noise figure ($NF$) & 5 dB\\
\hline
% \hline
% SSFM step size [km] & 10 & 0.1 & /\\
% \hline
% Number of channels & 3 & \multicolumn{2}{c|}{5}\\
% \hline
% Samples per symbol & 8 & 16 &\\
\end{tabular}
\end{table}

Two stage optimization is performed: pre-training and training. The pre-training is performed on a simplified channel model in order to increase the computation speed. Final training is then applied for convergence on the desired channel. For the pre-training stage, a shorter link of $5$ spans, 8 samples per symbol, 3 WDM channels and a large SSFM step size of 10~km is used. The optimization in this stage is done for the launch power per channel $P_s=0.5$~dB, which was found to be the optimal launch power for a non-shaped QAM. Afterward, a few epochs of training are performed on the desired link of 10 spans, 16 samples per symbol, 5 WDM channels and a SSFM step size of 100 m for convergence. In this stage, the AE is trained separately for each launch power and later on it is evaluated on the same launch power. Minor improvements may be expected by optimizing on each of the considered launch powers already in the pre-training stage, however, at the expense of significantly increasing computational time. Alternatively, the launch power may potentially be added to the optimization process as in \cite{Jones2018}, which is left for future research.

In the training stage, when BP algorithm is applied, checkpointing from \cite{GaiarinJLT} is employed at every 10 steps, corresponding to every 1 km. Checkpoining is a memory saving method when training very deep NNs in which only certain \emph{checkpoints} are saved instead of saving the full computational graph \cite{chen2016training}. This comes at a price of computational speed because the parts of the graph that are not saved are re-computed during BP. 

As indicated before, the AE is trained to optimize the performance of one of the polarizations of the central channel. The AE architecture is the same as in \cite{Ognjen_JLT, Ognjen_ECOC_Extension} and it is given in Table \ref{tab:MI_Setup}. The weight set $\mathbf{w}$ is initialized using Glorot initialization \cite{Glorot2010}. In each training epoch, a new sample set of size $N=256 \cdot M$ is generated with uniformly distributed one-hot encoded vectors and divided into batches of size $B=32 \cdot M$. The Adam optimizer \cite{kingma2014adam} with learning rate optimized to $0.001$ was used as the BP algorithm. The RL-based optimization also relies on the Adam optimizer with the same learning rate and the policy variance is $0.01$. In this case, a single epoch consists of $20$ training iterations of the encoder and $20$ iterations of the decoder. The hyperparameters of the CKF algorithm are $Q=10^{-8}$, $R=10^{-6}$ and the initial covariance of the weights is $\mathbf{P} = 10^{-4}\mathbf{I}$, where $\mathbf{I}$ is an identity matrix. Details on these parameters are given in \cite{Ognjen_JLT}. It should be mentioned that the hyperparameters of the three optimization algorithms are chosen as a result of a coarse optimization.

\begin{table}[!t]
\renewcommand{\arraystretch}{1.3}
\caption{Parameters of the encoder and decoder neural network for MI optimization.}
\label{tab:MI_Setup}
\centering
\begin{tabular}{|c||c|c|}
\hline
 & Encoder NN & Decoder NN\\
\hline
Number of input nodes & $M$ & 2\\
\hline
Number of hidden layers & 0 & 1\\
\hline
Number of nodes per hidden layer & 0 & $M/2$\\
\hline
Number of output nodes & 2 & $M$\\
\hline
Bias & No & Yes\\
\hline
Hidden layer activation function & None & Leaky Relu\\
\hline
Output layer activation function & Linear & Softmax\\
\hline
Cost function & \multicolumn{2}{c|}{Categorical cross-entropy}\\
\hline
\end{tabular}
\end{table}

\subsection{Impact of quantization on geometric constellation shaping} \label{ssec:Quantization}

One of the main drawbacks of GCS identified by the community is the potentially higher required quantization due to the irregular position of the points on each I and Q dimension \cite{Qu2019}. The non-quantization component of the digital-to-analog converter (DAC) and analog-to-digital converter (ADC) noise is frequency dependent \cite{Laperle:14} and may be considered similar for both GCS and uniform QAM. Then, in order to gauge the effect and requirements of the quantization noise in particular, a relatively low DAC/ADC resolution is studied.

\begin{figure}[htb]
\centering
\includegraphics[width=\columnwidth]{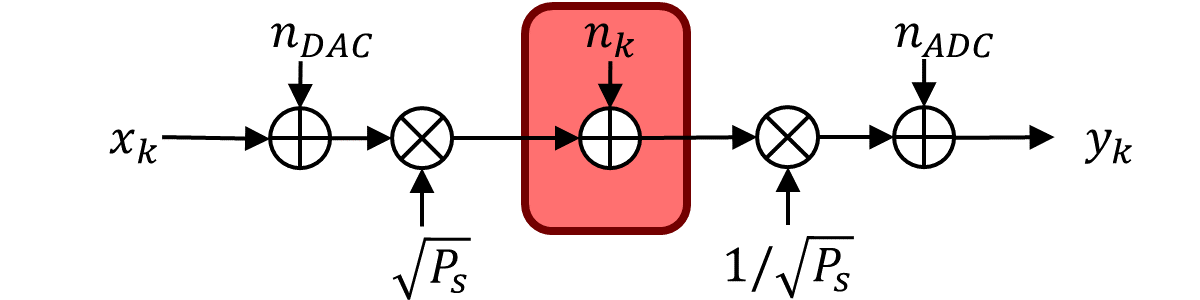}
\caption{Fiber-optic channel modeled by NLIN which includes DAC and ADC quantization noise.}
\label{fig:quantization_channel}
\end{figure}

For this analysis, the fiber-optic channel was modeled using a simpler nonlinear interference noise (NLIN) \cite{Dar2014} channel model which operates at one sample per symbol. This model takes into account the nonlinear interference dependence on the launch power per channel $P_s$ and the moments of the constellation. Since it is a one sample per symbol channel model, pulse shaping is not applied. In the NLIN channel model, the nonlinear effects degrading the transmitted signal are modeled as additive Gaussian noise with a variance $\sigma_{NLIN}^2(P_{s},\mu_4,\mu_6)$ that is determined by the parameters of the fiber communication channel. The high order moments are defined as
\begin{equation}
    \mu_4 = \frac{\mathbb{E}[|X|^4]}{(\mathbb{E}[|X|^2])^2} \quad \text{and} \quad
    \mu_6 = \frac{\mathbb{E}[|X|^6]}{(\mathbb{E}[|X|^2])^3}.
\end{equation} 
The total noise corrupting the signal is expressed as
\begin{equation}
    \sigma_n^2 = \sigma_{ASE}^2 + \sigma_{NLIN}^2(P_{s},\mu_4,\mu_6) \text{,}
    \label{eq:noise}
\end{equation}
where $\sigma_{ASE}^2$ is the variance of the amplified spontaneous emission (ASE) noise. The parameters of the channel model are the same as in the previous section, given in Table~\ref{tab:channel}. The NLIN model including quantization noise resulting from DAC and ADC are embedded into an AE as shown in Fig. \ref{fig:quantization_channel}. The quantization noise follows a uniform distribution determined by the number of quantization bits and the optimized dynamic range of the DAC and ADC \cite{Gray98, Proakis_DSP},
\begin{equation}
    n_{DAC/ADC} \sim \mathcal{U} \bigg(-\frac{A_{peak}}{2^{N_{bits}-1}},\frac{A_{peak}}{2^{N_{bits}-1}}\bigg),
    \label{eq:quantization_noise}
\end{equation}
where $N_{bits}$ is the number of quantization bits. Here, the peak amplitude $\mathbf{A}_{peak}$ of the transmitted signal determines the dynamic range of the two converters. The peak amplitude of the signal is chosen such that $\mathbf{A}_{peak}=1.2 \cdot [\mathrm{max}_i \mathrm{Re}[x^{(i)}], \mathrm{max}_i \mathrm{Im}[x^{(i)}]]$, where $(i)$ indicates the $i$-th constellation point. The factor $1.2$ is chosen to accommodate for power fluctuations and pulse shaping. It should be mentioned that after the DAC the signal is linearly modulated ($\sqrt{P_{s}}$ re-scaling) and that there is linear coherent detection ($1/\sqrt{P_{s}}$ re-scaling) before the ADC.
%For simplicity, flat frequency responses of the DAC and ADC were considered. It should be mentioned that the power rescaling by the launch power $P_{s}$ is done after adding the DAC quantization noise and rescaled again by the reciprocal of the launch power $1/P_{s}$ before adding the ADC quantization noise.

\begin{table}[htb]
\caption{Parameters of the encoder and decoder neural network for GMI optimization.}
\label{tab:GMI_Setup}
\centering
\begin{tabular}{|c||c|c|}
\hline
 & Encoder NN & Decoder NN\\
\hline
Number of input nodes & $m$ & 2\\
\hline
Number of hidden layers & 4 & 4\\
\hline
Number of nodes per hidden layer & 256 & 256\\
\hline
Number of output nodes & 2 & $m$\\
\hline
Bias & Yes & Yes\\
\hline
Hidden layer activation function & Relu & Relu\\
\hline
Output layer activation function & Linear & Sigmoid\\
\hline
Cost function & \multicolumn{2}{c|}{Binary cross-entropy}\\
\hline
\end{tabular}
\end{table}

For this channel model, optimizations of the AE with respect to both categorical CE and LL cost functions are done in order to maximize MI and GMI, respectively. For both performance metrics, the optimization of the AE is done through the Adam optimizer with learning rate $0.001$. The AE architecture when optimizing with respect to GMI is given in Table \ref{tab:GMI_Setup}.

When optimizing with respect to the categorical CE cost function, the sample and batch sizes are the same as in Section \ref{ssec:SSFM}. However, when optimizing with respect to LL, this training procedure often converges to local minima, which was also shown in \cite{Gumus2020}. To avoid converging to local minima, a different training procedure is applied and it relies on using an initial small batch size to have a more stochastic gradient estimation. This training procedure was inspired by what was implicitly done in \cite{Jones2019,rode2021geometric}. In each training epoch, an identical sample set of size $N=32 \cdot M$ is used when optimizing with respect to LL. In this set, the number of times each of the symbols, i.e. each combination of $m$ bits, occur is fixed to 32. In this case, the batch size $B$ varies throughout the training and it starts with the smallest size possible in which every $m$-bit combination occurs exactly once, i.e. $B=M$. When the optimization reaches a temporary convergence with the given batch size, the batch size is doubled. This procedure continues until the batch size reaches $B=N$ and when the optimization converges for this batch size, an early stop criterion terminates the optimization. Throughout the training procedure, the number of batches used to divide the sample set decreases. This training procedure will be referred to as the adaptive batch size training and a comparison with a fixed batch size (non-adaptive) is provided. After the training, the learned bit-labels are stored as look-up tables (LUTs) which are then applied during the testing.% It should be emphasized that the deterministic sample set and the order of the symbols is not important because the used channel does not have memory.

\begin{figure}[!t]
\centering
\includegraphics[width=\columnwidth]{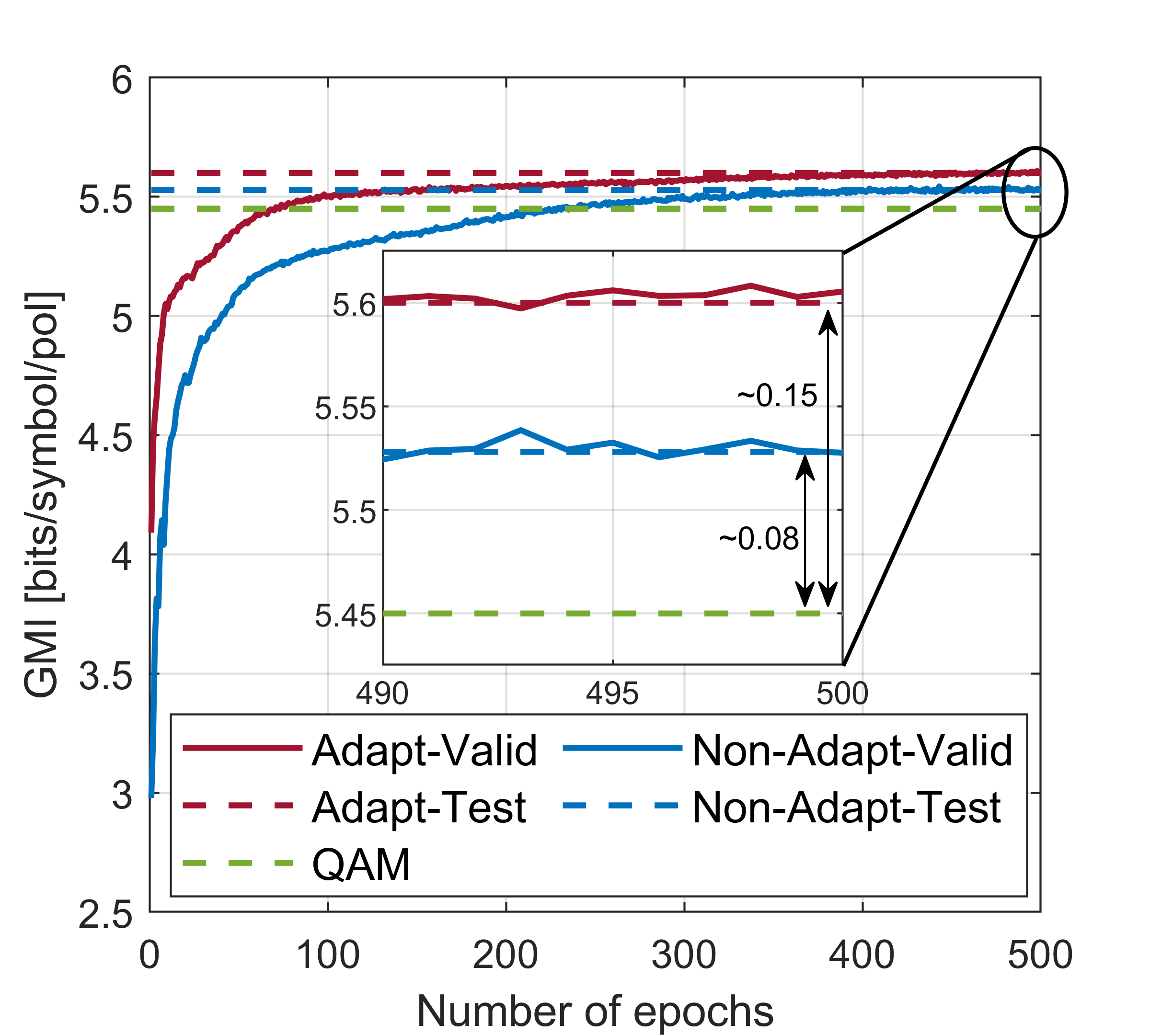}
\caption{Evolution of generalized mutual information with respect to the number of epochs for adaptive and non--adaptive batch size training. The results of the validation are represented with solid lines, whereas the results of the final tests  with dashed lines.}
\label{fig:Batch_adaptation}
\end{figure}

The comparison between the adaptive and the non-adaptive batch size training is shown in Fig.~\ref{fig:Batch_adaptation} by demonstrating the evolution of the GMI with respect to the number of epochs. This training was performed on the channel model shown in Fig. \ref{fig:quantization_channel} for an 8 bit quantization and constellation size $M=256$. The presented GMI results were obtained through validation at each training epoch. The GMI performance of 256QAM is included to provide a comparison in the achieved GCS gain. In the case of the non-adaptive batch size, the sample set is generated and divided into batches in the same way as in the optimization with respect to MI. The shaping gain in this case is only around $\sim 0.08$~bits/symbol which implies that the optimization might have converged to a local minimum. The validation results of the adaptive batch size training demonstrates faster convergence and a superior GMI performance compared to the non-adaptive training procedure. In this case, the achieved gain compared to 256QAM is almost double the gain which was achieved with the non-adaptive batch size. It should be mentioned that even though a higher GMI is achieved, it is still possible that the adaptive training converged to local minima.

\section{Numerical results} \label{results}

\subsection{Comparison of optimization algorithms}

The considered size of the constellation is $M=64$ and the presented results are acquired during testing which was done by averaging $10$ simulations with $10^5$ symbols per simulation in each case. A square QAM and the AWGN channel-optimized iterative polar modulation (IPM) \cite{Djordjevic2010} are used as the benchmark in this study. %\textcolor{red}{The AE is trained separately for each of the launch power and it is evaluated on the same launch power it was trained on.}

\begin{figure}[!t]
\centering
\includegraphics[width=\columnwidth]{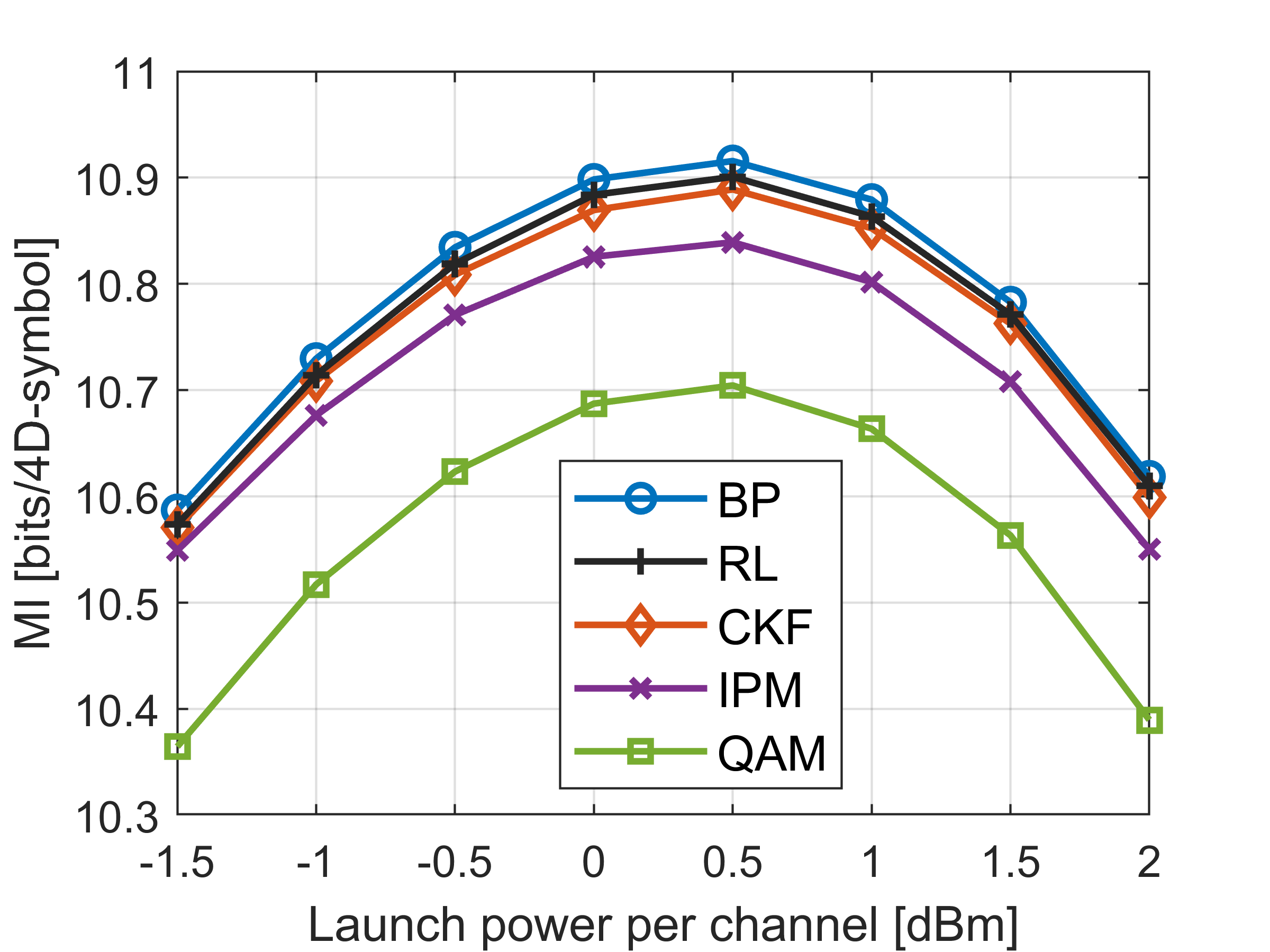}
\caption{Mutual information with respect to launch power for QAM, IPM, and constellations optimized with BP, RL and CKF. Comparison of optimization algorithms on a SSFM 1000km dual polarization 5 channel WDM system for constellation size $M=64$.}
\label{fig:Algorithm_comparison}
\end{figure}

In Fig. \ref{fig:Algorithm_comparison}, the MI performance of the QAM, IPM and constellations optimized with BP, RL and CKF algorithms are shown. In order to resemble a more practical coded modulation scheme, the presented MI was acquired using a mismatched Gaussian receiver. The decoder NN is used only to facilitate training. The results show that the BP algorithm learns the best performing constellation with a gain of $0.21$~bits/4D-symbol with respect to QAM. The constellations learned with RL and CKF have small penalties compared to the constellation learned with BP of around $0.015$ and $0.027$ bits/4D-symbol, respectively.

\begin{figure}[!t]
\centering
\subfloat[]{\includegraphics[width=\columnwidth]{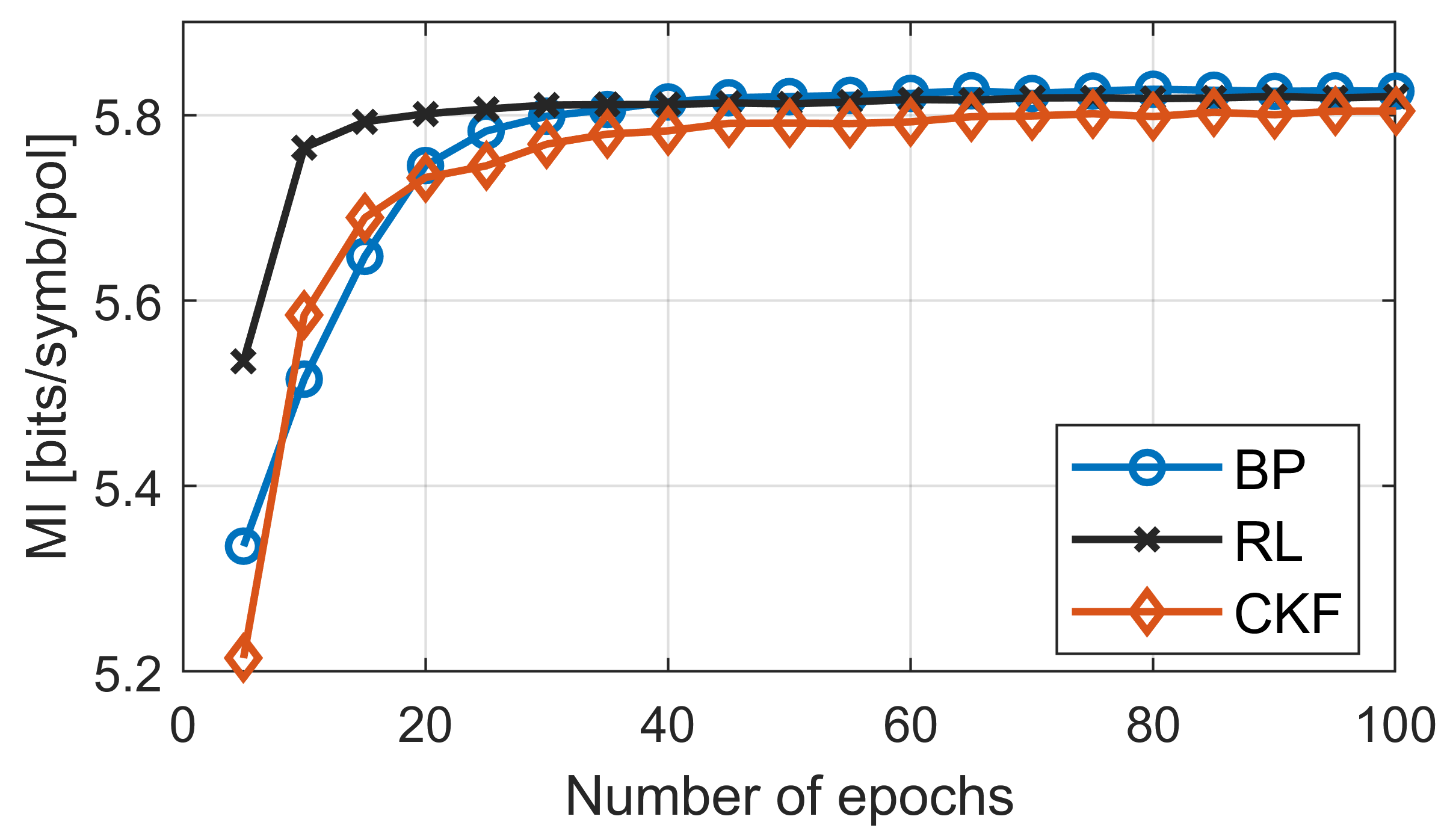}}
\hfil
\subfloat[]{\includegraphics[width=\columnwidth]{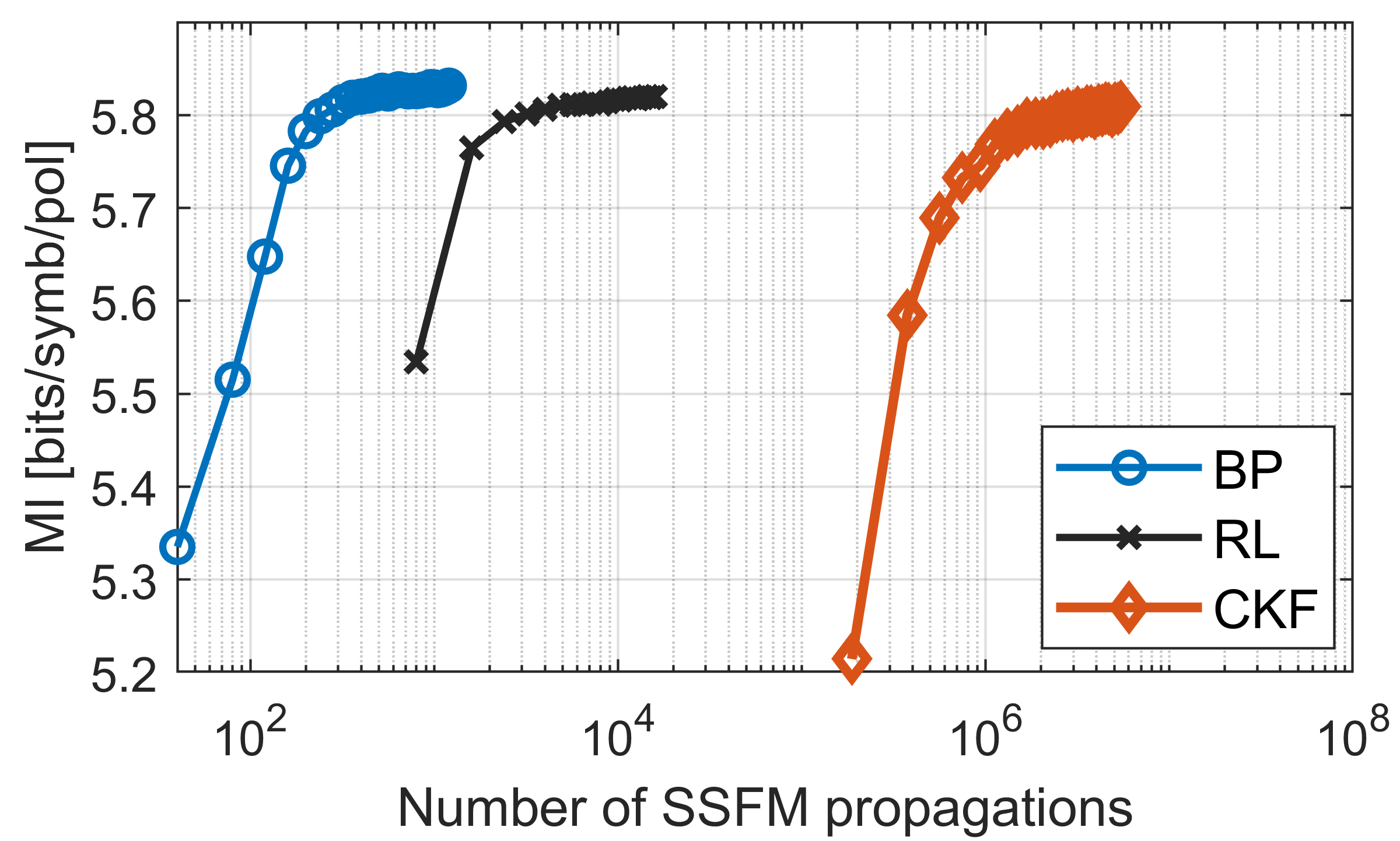}}
\caption{Mutual information convergence in the pre--training stage with respect to: (a) number of epochs and (b) number of SSFM propagations for BP, RL and CKF.}
\label{fig:SSFM_convegence}
\end{figure}

In Fig. \ref{fig:SSFM_convegence}, the evolution of the MI performance with respect to: (a) number of epochs and (b) number of SSFM propagations during the pre-training is shown. These results are obtained by validating the encoder (constellation) on every fifth epoch using the mismatched Gaussian receiver. The main interest is in the convergence of the MI obtained by the mismatched Gaussian receiver because it was used for testing. Here, the convergence will be defined as $99.5\%$ of the final MI. Observing Fig. \ref{fig:SSFM_convegence}~(a), the required  number of epochs for convergence is 30, 20 and 35 for BP, RL and CKF, respectively, which are of similar magnitude. However, this observation may be misleading because it does not show the true convergence and complexity of the three algorithms due to their different optimization approaches. Instead, convergence with respect to number of channel propagations is observed in Fig. \ref{fig:SSFM_convegence}~(b). From this figure, it can be observed that for the RL and the CKF algorithms, the number of SSFM propagations is drastically higher than for the BP algorithm. The RL algorithm requires an order of magnitude more SSFM propagations compared to BP, whereas the CKF algorithm requires three orders of magnitude more. A potential mitigation of the CKF complexity is the possibility to perform SSFM computations in parallel. It should be noted that for the BP algorithm, the recalculation due to checkpointing is not taken into consideration for this analysis.

\begin{table}[htb]
 \renewcommand{\arraystretch}{1.5}
 \caption{Comparison between AE training algorithms.}
 \label{tbl:comparison}
 \centering
 \begin{tabular}{|c||c|c|c|}
 \hline
 & BP & RL & CKF \\
 \hline
 \# of SSFM props. per batch & 1 & 20 & $2*N_w \approx 4500 $ \\
% props. per batch &   &   & $\approx4500$ \\
%  \hline
%  \# of epochs  & 45 & 30 & 40 \\
%  for conversion &   &   &   \\
 \hline
 Processing & serial & serial & parallel \\
 \hline
 Tx and Rx optimization & joint & iterative & joint \\
 \hline
 Usable in experimental test-bed & no & yes & yes \\
 \hline
 Require memory for gradient & yes & no & no \\
 %for gradient &   &   &   \\
 \hline
 \end{tabular}
\end{table}

The properties of the three algorithms are summarized in Table~\ref{tbl:comparison}. The constellations trained using BP demonstrates the best performance in terms of MI and requires the least amount of SSFM propagations. However, in order to perform an optimization using the BP algorithm, the embedded channel model has to be differentiable. In addition, this approach requires a significant amount of memory allocation because each SSFM step is stored as a part of the computational graph used for gradient calculation. The RL and CKF optimization methods do not require a differentiable channel model, therefore they could be potentially used in an experimental test-bed. However, due to the number of required channel propagations this could prove also to be challenging.

\subsection{Impact of quantization on geometric constellation shaping}

\begin{figure}[!t]
\centering
\subfloat[]{\includegraphics[width=\columnwidth]{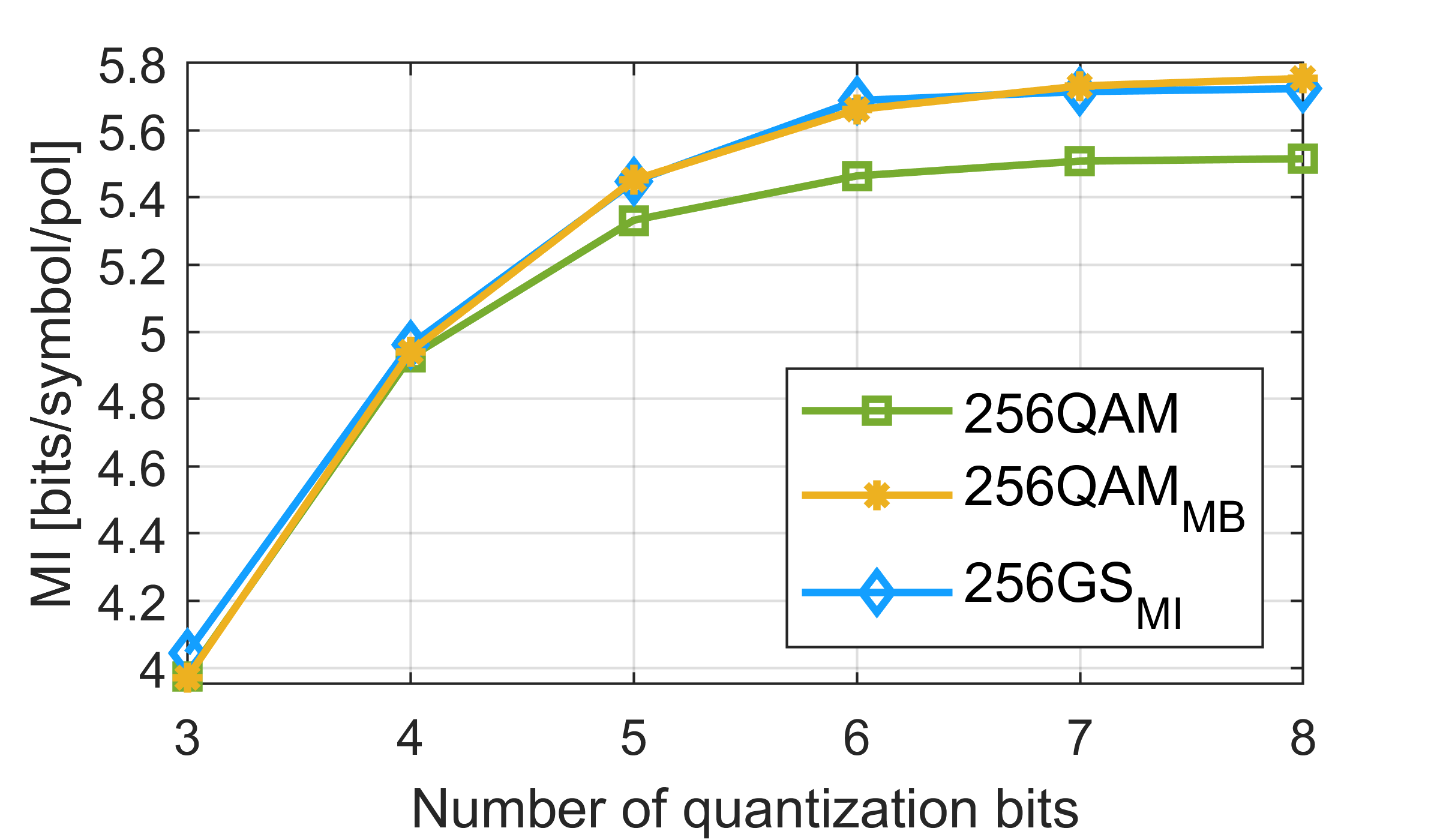}}
\hfil
\subfloat[]{\includegraphics[width=\columnwidth]{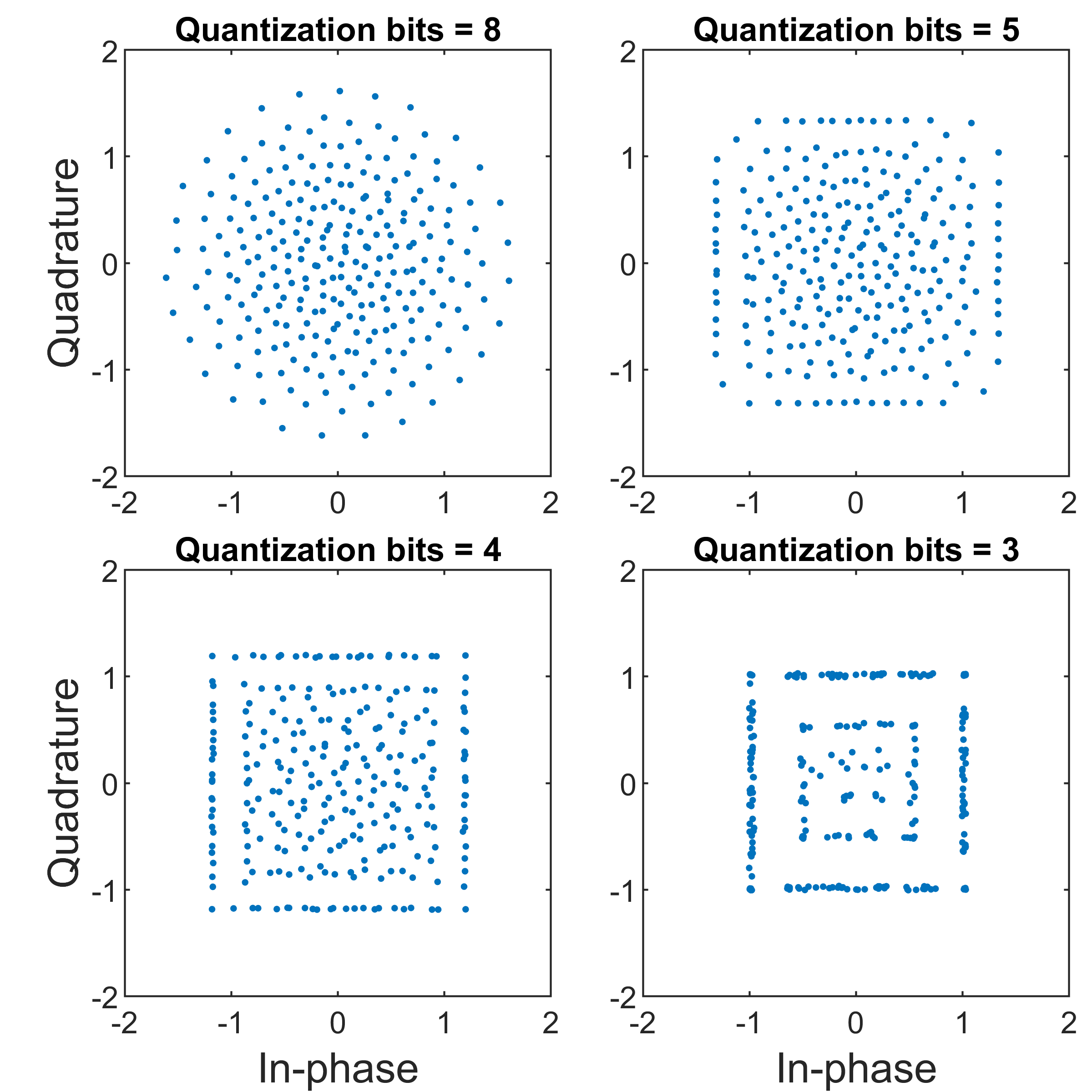}}
\caption{(a) Performance in mutual information with respect to number of quantization bits for 256QAM, $\text{256QAM}_{\text{MB}}$ and $\text{256GS}_{\text{MI}}$. (b) Constellations learned for quantization of $8$, $5$, $4$ and $3$ bits.}
\label{fig:MI_quantization}
\end{figure}
%\textcolor{red}{One of the main drawbacks of GCS identified by the community is the potentially higher required quantization due to the irregular position of the points on each I and Q dimension \cite{Qu2019}. The non-quantization noise of DAC and ADC is frequency dependent \cite{Laperle:14} and may be considered similar for both GCS and uniform QAM. Then, in order to gauge the effect and requirements of the quantization noise in particular, a relatively low DAC/ADC resolution is studied.} 

The considered constellation size is $M=256$ and the considered number of quantization bits are given by the set $\{3,4,\dots,8\}$. Each of the trained AEs was tested with the same channel parameters as the ones used for training. The testing was done by running $10$ simulations with $10^5$ symbols per simulation in each case and the presented results are the average over those simulations. As benchmarks, the Gray-coded uniform and probabilistic shaped QAM constellations are studied. For probabilistic shaping, the Maxwell-Boltzmann (MB) probability mass function \cite{buchali2016rate} optimized for each number of quantization bits is used, which is indicated as e.g. {$\text{256QAM}_{\text{MB}}$}.

The \emph{MI performances} as a function of number of quantization bits of 256QAM, $\text{256QAM}_{\text{MB}}$ and geometrically shaped constellations \emph{optimized with respect to MI} (denoted as $\text{256GS}_{\text{MI}}$) are shown in Fig. \ref{fig:MI_quantization}~(a). The GCS gain compared to QAM is the highest at 8 bit quantization and it amounts to $\sim 0.2$~bits/symbol. With the reduction of quantization bits, the shaping gain deteriorates to $\sim 0.15$~bits/symbol at 5 bit quantization. For 4 and 3 quantization bits, the GCS gain falls below $0.05$~bits/symbol and the achieved MI is similar to the MI performance of QAM. The performance obtained with GCS, $\text{256GS}_{\text{MI}}$, is similar to the one obtained with PCS, $\text{256QAM}_{\text{MB}}$. In Fig. \ref{fig:MI_quantization}~(b), the constellations learned for quantization of 8, 5, 4 and 3 bits are shown. When the number of quantization bits is higher, the dominant noise source is the NLIN and in that case the learned constellation has a circular form. As the number of quantization bits decreases, the quantization noise becomes the dominant noise source and in this case the learned constellation has a rectangular form. It can be observed that in the severe quantization case, the learned constellation has only a few distinct levels and points in an effort to reduce the quantization noise impact.

% \begin{figure*}[htb]
% \centering
% \includegraphics[width=\linewidth]{Figures/AIR_GMI_MI_with64QAM.eps}
% \caption{Performance in achievable information rate with respect to the number of quantization bits for 64QAM, 256QAM, 256GS optimized with respect to MI, and 256GS optimized with respect to GMI. The GMI performance is represented with solid lines and the MI performance with dashed lines.}
% \label{fig:AIR_GMI_MI}
% \end{figure*}

\begin{figure}[!t]
\centering
\subfloat[]{\includegraphics[width=\columnwidth]{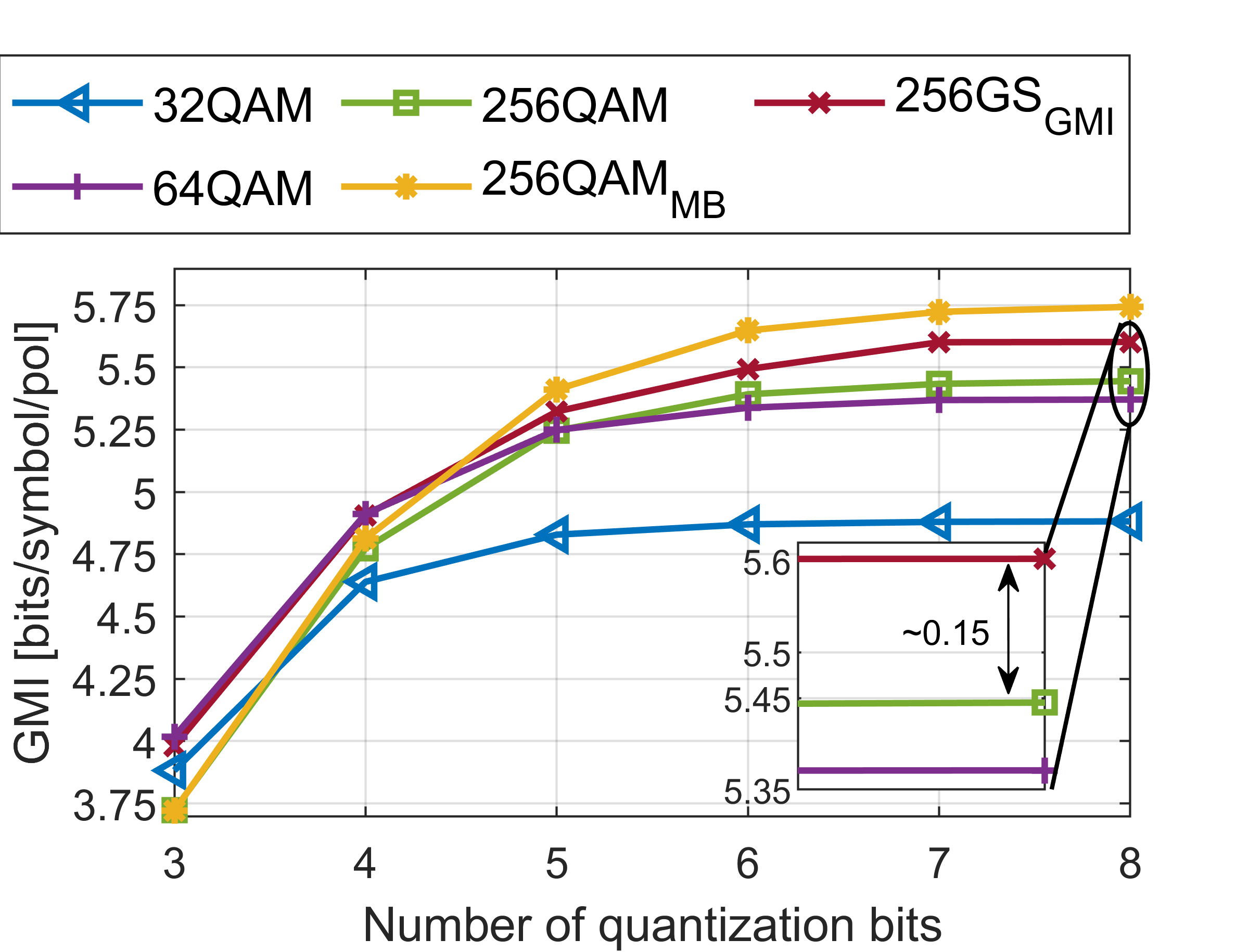}}
\hfil
\subfloat[]{\includegraphics[width=\columnwidth]{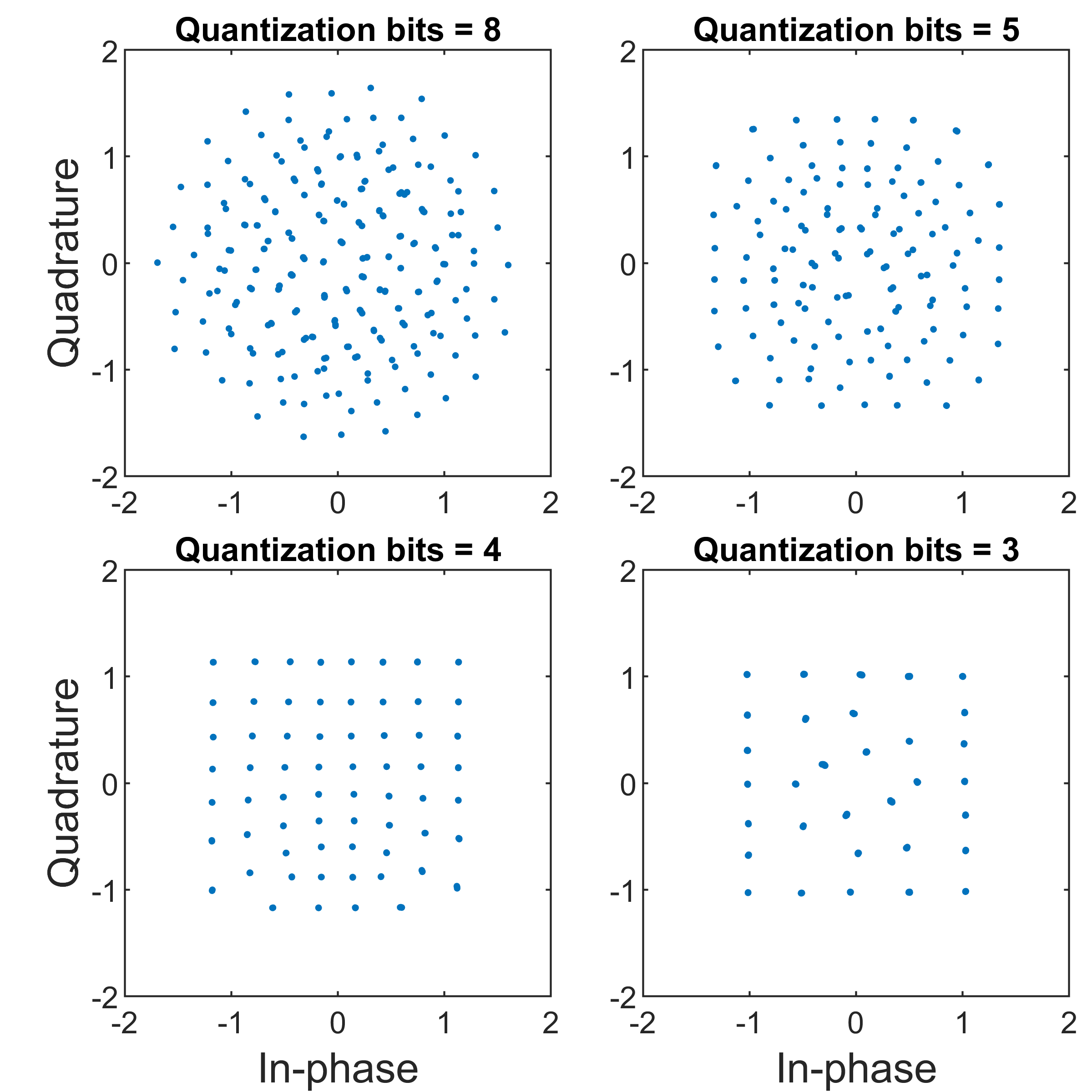}}
\caption{(a) Performance in generalized mutual information with respect to number of quantization bits for 32QAM, 64QAM, 256QAM, $\text{256QAM}_{\text{MB}}$ and $\text{256GS}_{\text{GMI}}$. (b) Constellations learned for quantization of $8$, $5$, $4$ and $3$ bits.}
\label{fig:GMI_quantization}
\end{figure}

The \emph{GMI performances} as a function of number of quantization bits of 32QAM, 64QAM, 256QAM, $\text{256QAM}_{\text{MB}}$ and geometrically shaped constellation \emph{optimized with respect to GMI} (denoted as $\text{256GS}_{\text{GMI}}$) are shown in Fig.~\ref{fig:GMI_quantization}~(a). The shaping gain compared to 256QAM for 8 bit quantization of the constellation optimized with respect to GMI is $0.15$~bits/symbol which is lower than the $0.2$~bits/symbol gain shown when observing and optimizing with respect to MI. The shaping gain compared to 256QAM degrades as the number of quantization bits decreases from 8 to 5, however, the gain improves for 4 and 3 bit quantization. The highest achieved GSC gain compared to 256QAM is $\sim 0.25$~bits/symbol and it is achieved for 3 bit quantization. In this regime, the optimal QAM size is reduced to $M=64$, and achieves similar GMI to the $\text{256GS}_{\text{GMI}}$ constellation. Whereas, compared to a QAM constellation of size $M=32$, the $\text{256GS}_{\text{GMI}}$ constellation achieves gain of around $0.1$~bits/symbol. Also, due to lack of ideal Gray coding, the GMI performance of 32QAM is penalized compared to 64QAM. When observing GMI, the gains of PCS are higher than the ones obtained with GCS with a difference of up to $0.15$~bits/symbol. However, PCS has a higher complexity compared to GCS because it requires a distribution matcher and dematcher. Similar to uniform QAM, for severe quantization, $\text{256QAM}_{\text{MB}}$ is penalized, and the modulation size would need to be reduced to maintain shaping gain. In Fig. \ref{fig:GMI_quantization}~(b), the constellations learned for quantization of 8, 5, 4 and 3 bits are shown. The shape of the learned constellations optimized with respect to GMI are similar to the ones optimized with respect to MI. For a higher number of quantization bits, the shape of the constellation is circular and as the number of quantization bits decreases, the learned constellations take a more rectangular form of apparent lower cardinality. The rectangular form of the constellation decreases the effect of quantization noise, while the decrease of the cardinality makes the constellation more robust to highly noisy environments. Observe, for systems dominated by uniformly distributed quantizaition noise, uniform points position are optimal \cite{Gray98}.

\begin{figure}[htb]
\centering
\includegraphics[width=0.95\columnwidth]{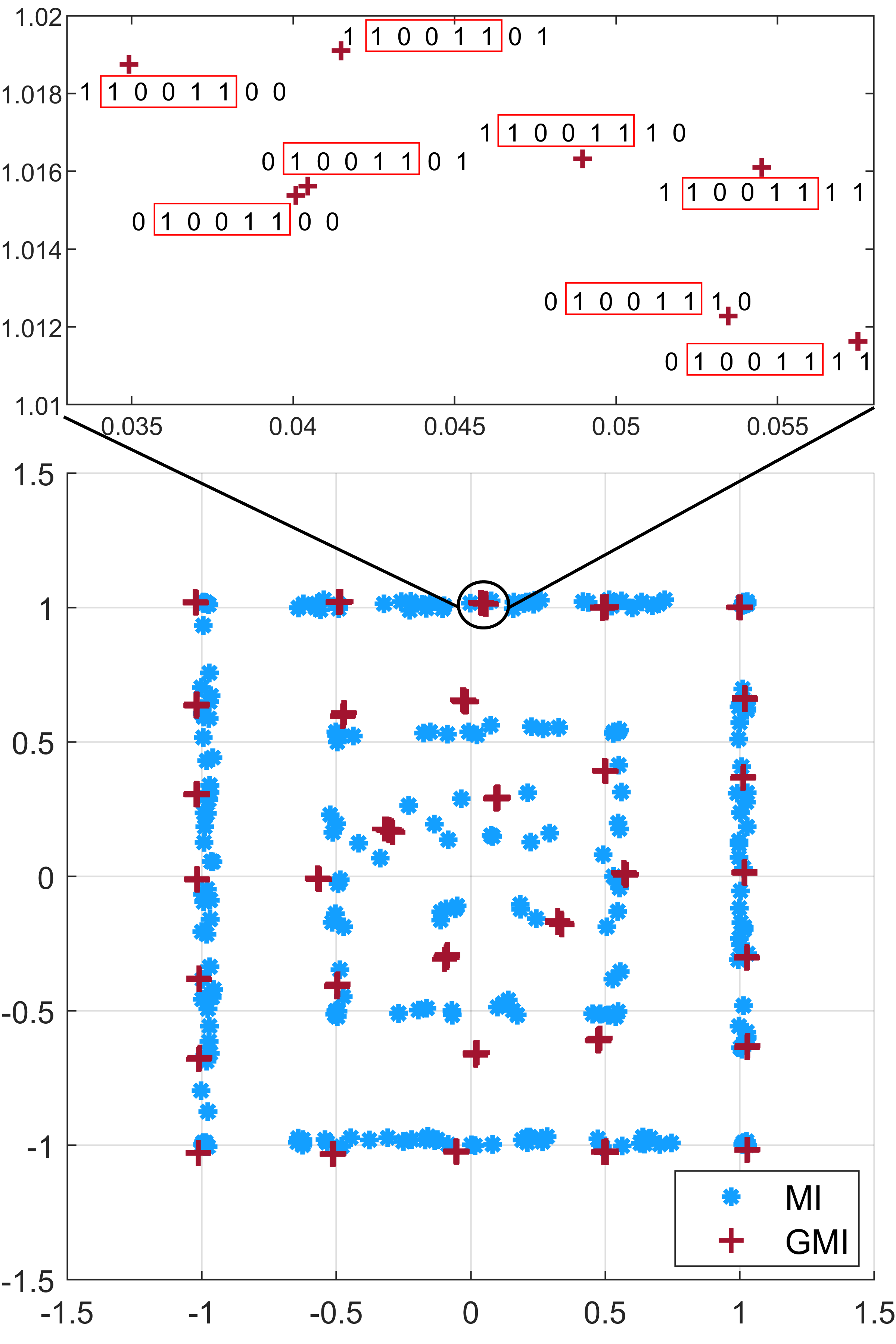}
\caption{Constellations optimized with respect to MI and GMI for $3$ bit quantization. A point of the GMI optimized constellation is zoomed in to show that eight points have collapsed to the same location. The labeling of these eight points is included and the five bits that are the same for all eight points are annotated with a red rectangular box.}
\label{fig:MI_GMI_constellations}
\end{figure}

This effect is further analyzed in Fig. \ref{fig:MI_GMI_constellations}, where the constellations optimized with respect to MI and GMI for 3 bit quantization are shown. The constellation optimized with respect to MI forms two squares with amplitudes of $\sim 1$ and $\sim 0.5$ and with some points inside the inner square. Most of the points in this constellation are close to each other but still distinguishable. The fact that the points with low Euclidean distance are not merged together does not impact the achieved MI because similar MI performances can be achieved around the optimum parameters. However, this is not the case when optimizing for GMI. When optimizing for GMI, the optimizer merges points that are close to each other and assigns them labels with low Hamming distance. A zoomed in version of one of these points is shown and it can be noticed that eight symbols are positioned in an interval of $\sim 0.02$. All of these eight symbols have the same five bits in their labels which is indicated with a red rectangular box, whereas the rest of the bits can have any combination. In this case, the AE effectively acts as a compressor, optimal for the target SNR and achievable rate conditions. It should be mentioned that for different models of quantization noise, the learned constellations and achieved performances may differ. However, our proposed method is still applicable to these other models, as long as they are differentiable.
%This effect is further analyzed in Fig. \ref{fig:MI_GMI_constellations}, where the constellations optimized with respect to MI and GMI for 3 bit quantization are shown. The constellation optimized with respect to MI forms two squares with amplitudes of $\sim 1$ and $\sim 0.5$ and with some points inside the inner square. Most of the points in this constellation are close to each other but still distinguishable. When optimizing for GMI, labeling of these points becomes difficult for the optimizer. Instead, the points are merged closer together, and assigned labels with low Hamming distance to each other. A zoomed in version of one of these points is shown and it can be noticed that eight symbols are positioned in an interval of $\sim 0.02$. All of these eight symbols have the same five bits in their labels which is indicated with a red box, whereas the rest of the bits can have any combination. In this case, the AE effectively acts as a compressor, optimal for the target SNR and achievable rate conditions.

\begin{figure}[t]
\centering
\includegraphics[width=\linewidth]{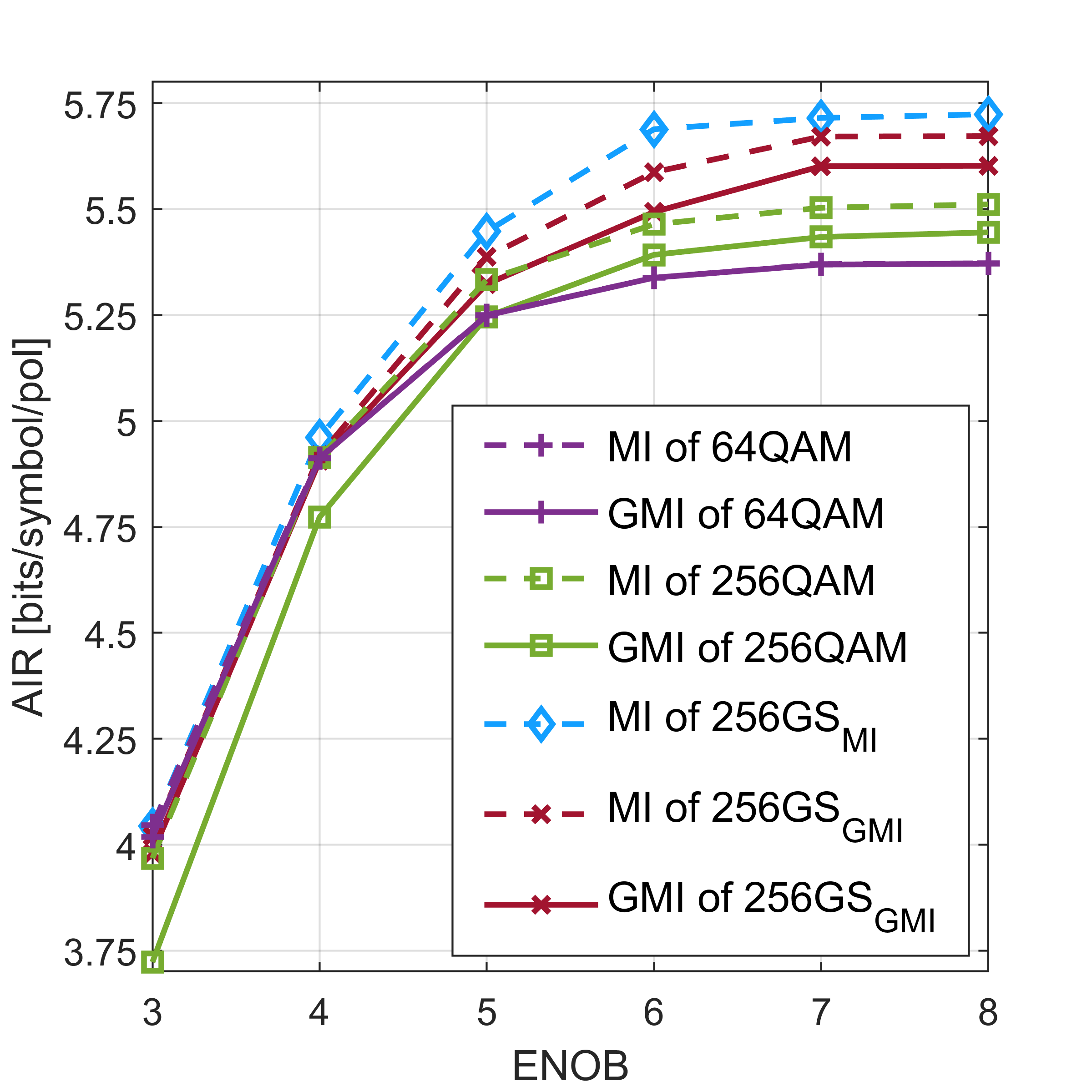}
\caption{Performance in achievable information rate with respect to the number of quantization bits for 64QAM, 256QAM, $\text{256GS}_{\text{MI}}$, and $\text{256GS}_{\text{GMI}}$. The GMI performance is represented with solid lines and the MI performance with dashed lines.}
\label{fig:AIR_GMI_MI}
\end{figure}

The AIR with respect to the number of quantization bits for the 64QAM, 256QAM, and 256GS optimized with respect to both MI ($\text{256GS}_{\text{MI}}$) and GMI ($\text{256GS}_{\text{GMI}}$) is shown in Fig.~\ref{fig:AIR_GMI_MI}. The probabilistic constellation shapes are excluded from this figure and analysis because the goal is only to observe the geometrically optimized constellation shapes. Both the MI and the GMI performances of the constellations are observed except for the constellation $\text{256GS}_{\text{MI}}$ for which only the MI performance is considered. In the case of $\text{256GS}_{\text{MI}}$, the optimal bit-labeling is not easily defined, therefore the GMI performance of this constellation was excluded.  The MI performance is represented with dashed lines, whereas the GMI performance with solid lines. When analysing the MI performance of the constellations, it can be noticed that the constellation $\text{256GS}_{\text{MI}}$ achieves the highest shaping gain compared to 256QAM for 8 quantization bits. The shaping gain in MI performance is slightly penalized in the case of $\text{256GS}_{\text{GMI}}$. For both optimizations, the shaping gain in MI performance degrades with the decrease of quantization bits. For quantization with less than 5 bits, the difference in MI performance between all constellations is marginal. The same cannot be observed when analysing the GMI performance of the constellations. In that case, the $\text{256GS}_{\text{GMI}}$ achieves shaping gain over the whole observed range of quantization bits for constellations of the same order. As it was already discussed, in the regime of a few quantization bits, the optimal QAM is reduced to $M=64$. These results imply that the optimizer can determine the required constellation size for the given channel conditions when the constellation is optimized with respect to GMI. %However, this is not the case when optimizing the constellation with respect to MI because for these lower SNR scenarios both constellations with 256 and 64 cardinality have similar MI performance.

\section{Conclusion} \label{Conclusion}
Several optimization algorithms were compared for autoencoder (AE)-based geometric constellation shaping (GCS) over the fiber-optic channel. The backpropagation (BP) algorithm is the obvious choice when the channel model is differentiable and fairly simple. For complex differentiable channels, memory saving methods such as checkpointing need to be applied in combination with BP. For non-differentiable and non-numerical channel models, the reinforcement learning-based and cubature Kalman filtering optimization methods present decent alternatives to BP with a slight performance degradation. However, the capability of optimizing over a non-differentiable channel comes at a price, as these two algorithms are quite computationally demanding and require significantly more channel propagations than BP.

Furthermore, the influence of digital-to-analog and analog-to-digital converter's bit resolutions on an AE-based GCS was analysed. The quantization effects of the converters were modeled by uniformly distributed noise sources. When optimizing the AE with respect to mutual information (MI), the shaping gain deteriorates for fewer quantization bits and at least a 5 bit resolution is required to have notable shaping gain. When optimizing the AE with respect to generalized MI (GMI), there is always shaping gain compared to standard square QAM constellation of the same order. The results imply that in this case, the optimizer overlaps the constellation points and adapts the constellation size to the channel conditions. The flexibility to vary the constellation size with a fine step while maintaining solid bit-metric performance allows the AE to achieve GMI gain over the entire range of quantization levels.

\section*{Acknowledgment}

This work was financially supported by the European Research Council through the ERC-CoG FRECOM project (grant agreement no. 771878), the Villum Young Investigator OPTIC-AI project (grant no. 29334), and DNRF SPOC, DNRF123.

\bibliographystyle{IEEEtran}
\bibliography{references.bib}
\end{document}